\documentclass{agujournal2019}
\usepackage{url}

\draftfalse

\journalname{JGR: Space Physics}

\usepackage{amsmath,amssymb,amsfonts,amstext,graphics,graphicx,txfonts,wrapfig}

\def\bq{\begin{equation}}
\def\eq{\end{equation}}
\def\bqy{\begin{eqnarray}}
\def\eqy{\end{eqnarray}}
\def\p{\partial}

\begin{document}

\title{Shear flow-interchange instability in nightside magnetotail causes auroral beads as a signature of substorm onset}
\authors{Jason Derr\affil{1}, Wendell Horton\affil{1}, Richard Wolf\affil{2}}
\affiliation{1}{Institute for Fusion Studies and Applied Research Laboratories,  University of Texas at Austin}
\affiliation{2}{Physics and Astronomy Department, Rice University}
\correspondingauthor{Jason Derr}{jderr@utexas.edu}

\begin{keypoints}
	\item A wedge model is extended to obtain a MHD magnetospheric wave equation for the near-Earth magnetotail, including velocity shear effects.
	\item Shear flow-interchange instability to replace shear flow-ballooning and interchange instabilities as substorm onset cause.
	\item WKB applicability conditions suggest nonlinear analysis is necessary and yields spatial scale of most unstable mode's variations.
\end{keypoints}

\begin{abstract}

A geometric wedge model of the near-earth nightside plasma sheet is used to derive a wave equation for low frequency shear flow-interchange waves which transmit $\vec{E} \times \vec{B}$ sheared zonal flows along magnetic flux tubes towards the ionosphere. Discrepancies with the wave equation result used in \citeA{KAL15} for shear flow-ballooning instability are discussed. The shear flow-interchange instability appears to be responsible for substorm onset. The wedge wave equation is used to compute rough expressions for dispersion relations and local growth rates in the midnight region of the nightside magnetotail where the instability develops, forming the auroral beads characteristic of geomagnetic substorm onset. Stability analysis for the shear flow-interchange modes demonstrates that nonlinear analysis is necessary for quantitatively accurate results and determines the spatial scale on which the instability varies.

\end{abstract}

\section{Introduction}

When the interplanetary magnetic field originating at the sun contains a southward magnetic field component, the solar wind causes magnetic reconnection on the dayside of the earth, followed by nightside reconnection in the magnetotail. The magnetic reconnection on the dayside changes the field configuration. Initially, there is a field line with both ends attached to the sun and a field line with both ends attached to the Earth. Reconnection then produces new field lines, one attached to the Earth's geomagnetic North Pole and extending into space and the other attached to the Earth's geomagnetic South Pole and extending into space, creating open flux tubes. The flow of plasma from the solar wind then produces an electric field which causes convection of the magnetic field lines towards the nightside of the Earth due to flux freezing. The open flux tubes which connect to the polar regions of the Earth thereby provide the duskward-directed electric field which drives the noon-midnight currents that carry the ionospheric ends of the magnetic field lines along with them. Reconnection on the nightside again closes magnetic field lines which connect to the Earth's geomagnetic North and South Poles, forming stretched closed flux tubes on the nightside of the Earth \cite{DUN61, KIV95}. A helpful diagram of the process is shown below in Figure \ref{Figure1}.

\begin{figure}[h!]
	\centering
	\includegraphics[width=0.6\textwidth]{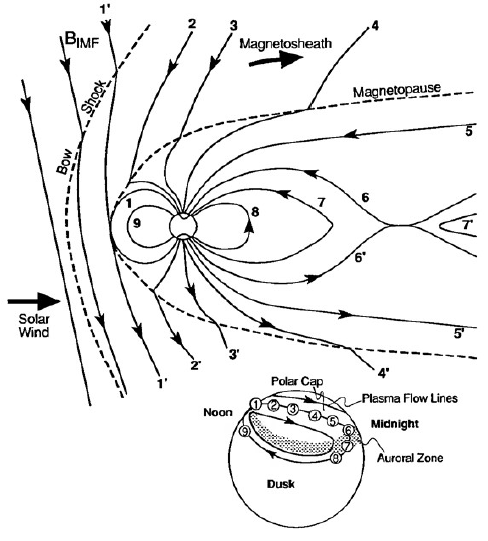}
	\caption{Solar wind-driven reconnection events for magnetic field lines. Field line numbers show the sequence of magnetic field line configurations. The first reconnection event occurs on the dayside of the Earth, where the southward interplanetary magnetic field 1' connects with the northward geomagnetic field line 1. The field lines are convected anti-sunward in configurations 2 and 2' through 5 and 5', reconnecting again as 6 and 6'. Then the substorm expansion phase is initiated and field lines connected to the geomagnetic North and South Poles dipolarize. Field line 7', now closed as a roughly teardrop-shaped plasmoid, then continues tailward into interplanetary space. The geomagnetic field line will then swing back around from midnight to noon to become field line 9. The inset below shows the positions of the ionospheric anchors of the numbered field lines in the northern high-latitude ionosphere and the corresponding plasma currents of the Earth's polar caps. The currents flow from noon to midnight, convecting the feet of the field lines nightward and then closing back at noon via a lower latitude field return flow. This figure is reproduced from \citeA{KIV95}.}
	\label{Figure1}
\end{figure}

Then a sequence of events referred to as a magnetic substorm occurs. The solar wind plasma deposits energy in the magnetotail during the growth phase of a magnetic substorm as the nightside magnetic field lines stretch tailward. This triggers a current disruption in the equatorial current sheet and initiates the expansion phase of the magnetic substorm, when the magnetic field lines snap back in response to the destabilization in the current sheet, and plasma is accelerated towards the polar regions of the earth's atmosphere. This sequence of events leads to the formation of aurorae in the E-layer of the Earth's ionosphere. Substorms occur about 5 times per day on average, and last for about 2-3 hours, but substorm onset occurs within a roughly 2 minute time span \cite{COP66,KIV95,WOLF95,ANG08-1,ANG08-2,ZEE04,ZOU10,MCP11,SER11,FOR14-1}.

At substorm onset (the initiation of the expansion phase), the most equatorward auroral arc suddenly brightens, followed by breakup of the arc and poleward expansion \cite{AKA64,DON08}. In the minutes leading to the breakup, small periodic fluctuations in the aurora aligned with magnetic longitude form \cite{NIS14}, seen below in Figure \ref{Figure2}. These fluctuations have come to be called ``auroral beads.'' Henceforth, ``longitudinal'' will be used to refer to magnetic longitude. Auroral beads have been found to be likely pervasive in onset arcs, and the exponential growth of the beads indicates that a plasma instability in the magnetosphere is responsible for substorm onset \cite{GALL14,KAL17}.

\begin{figure}[h!]
	\centering
	\includegraphics[width=0.7\textwidth]{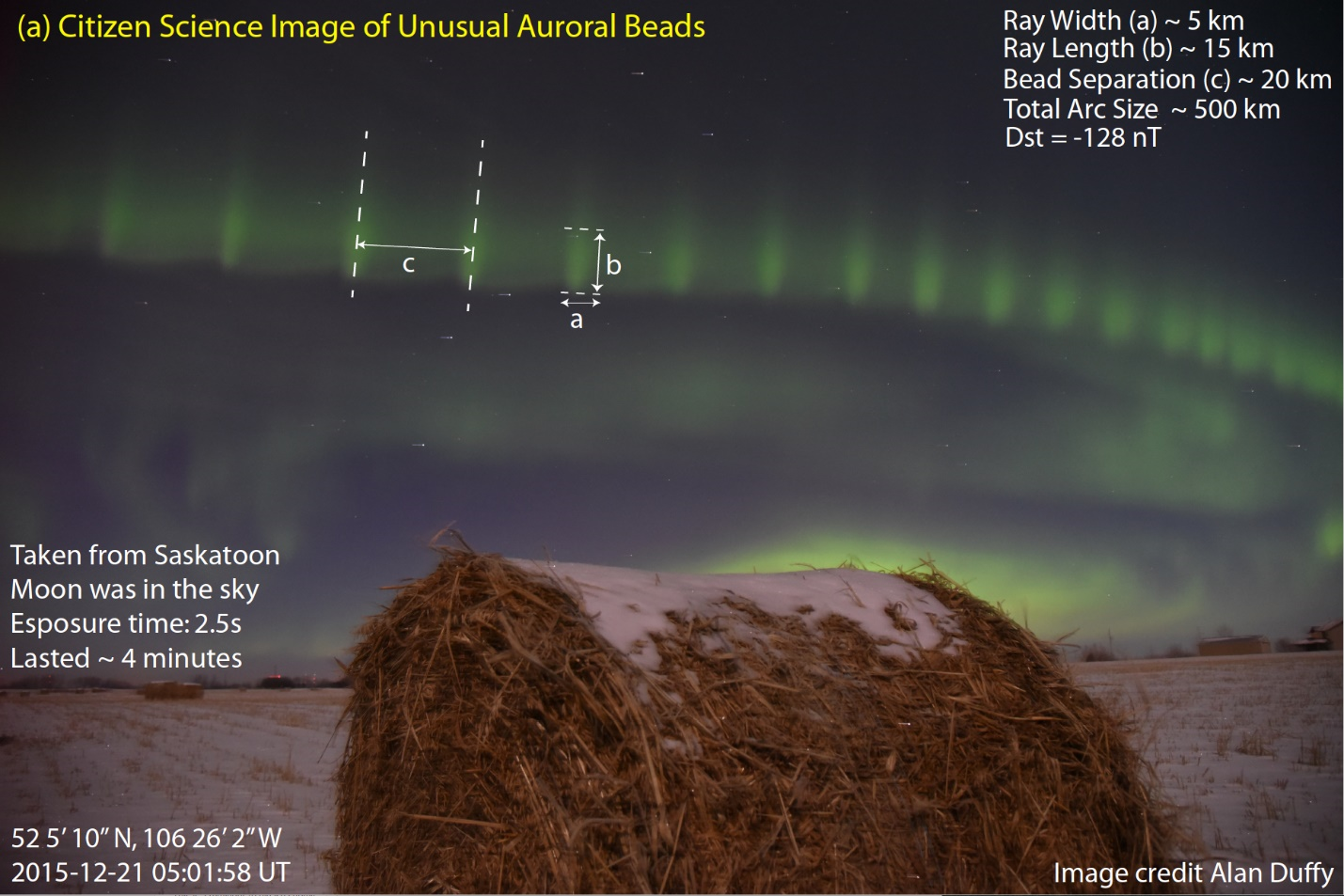}
	\caption{DSLR camera photograph of auroral beads above Canada, taken by Alan Duffy, Citizen Science. The bead spacings and other geometric information was ascertained by analysis of the star tracks.}
	\label{Figure2}
\end{figure}

\begin{figure}[h!]
	\centering
	\includegraphics[width=0.6\textwidth]{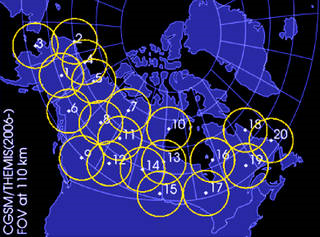}
	\caption{Distribution of the THEMIS All-Sky Imagers (ASIs), with Fields of View. \cite{NASA}}
	\label{Figure3}
\end{figure}

All-sky imagers (ASIs), which are a part of the NASA THEMIS mission to uncover the sequence of events which occur in the first few minutes of substorm onset, are distributed across North America, as seen in Figure \ref{Figure3}. They have a 1 km spatial resolution, and 3 s cadence image capturing capacity, and respond predominantly to 557.7 nm emissions. This spatio-temporal resolution is sufficient to capture the pertinent data for analyzing auroral bead structures for the green emissions corresponding to aurora at an altitude of approximately 110 km, namely the E layer \cite{MEN08,BUR08}.

\citeA{MOT12} used ASI data from auroral beads in the northern and southern hemispheres, and proposed a common magnetospheric driver. Ultra-low frequency waves occurring within minutes of substorm onset are observed in the magnetosphere at frequencies similar to those of the auroral beads, and a single event was analyzed by \citeA{RAE10} to demonstrate that the beading is characteristic of a near-earth magnetospheric instability triggering a current disruption in the central plasma sheet. Of the examined instabilities, cross-field current instability and shear flow-ballooning instability were the only two consistent with the analytical results. \citeA{KAL15} used the ASI data for 17 substorm events over a 12-hour time span throughout the auroral oval (pre-midnight sector) across Canada and Alaska to perform an optical-statistical analysis that yielded maximum growth rates for the beads as a function of longitudinal wavenumber, which were compared with theoretical calculations for growth rate dependence on wavenumber for various instabilities. Ultimately, the two mechanisms which remained unrefuted were the shear flow-ballooning instability and the cross-field current instability.

\begin{figure}[h!]
	\centering
	\includegraphics[width=0.7\textwidth]{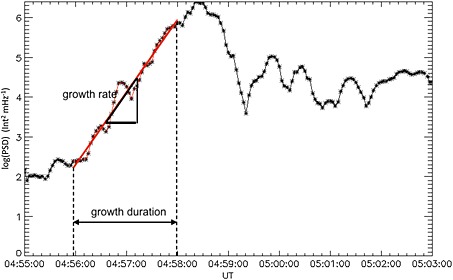}
	\caption{Exponential growth rate determination performed by \citeA{KAL15}. The log of the power spectral density plotted against time for a single wavenumber $k_y$ = 0.9*$10^4 \ m^{-1}$. This shows the duration of growth and growth rate from the linear fit. More details governing the fitting are given in \cite{KAL15}, from which this diagram is reproduced.}
	\label{Figure4}
\end{figure}

The statistical analysis involved first spatially Fourier transforming longitudinal keograms to obtain the power spectral density. The longitudinal wavenumbers $k_{y,E}$ measured in the ionosphere lay within the interval $k_{y,E}\in \left[ 0.5*10^{-4}\ m^{-1},\ 1.5*10^{-4}\ m^{-1} \right]$ during initial beading. The logarithm of the power spectral density was then plotted against time to determine the intervals of exponential growth for each wavenumber during onset. This is shown for one wavenumber in Figure \ref{Figure4}. Since the exponential growth of each mode had a unique well-defined growth-rate during the interval until the breakup, only one instability is operating to produce the growth for each event. The growth rates were then examined as a function of wavenumber for determination of the most unstable waves. The maximum growth rates lie in the range $\left[ 0.03\ Hz,\ 0.3\ Hz \right]$ with median growth rate $\gamma \sim 0.05\ Hz$. Note that wave propagation direction (eastward vs. westward) differed for the individual substorm events, but growth rates are independent of propagation direction \cite{NIS16}.

Subsequently, \citeA{KAL15} used the T96 model \cite{TSY95,TSY96,TSY96-M} to map the wavenumbers back to the equatorial magnetosphere to obtain the corresponding magnetospheric wavenumbers $k_y\in \left[ 2.5*{10}^{-6}\ m^{-1},3.75*{10}^{-6}\ m^{-1} \right]$, or wavelength interval ${\lambda }_{\bot }\in \left[ 1700\ km,\ 2500\ km \right]$. The T96 model underestimates field-line stretching (and spatial scales) during the substorm growth phase. Equilibrium magnetic field mapping is thus unreliable at substorm times. This implies that the T96 model will determine the location of the instability to be closer to Earth in the equatorial plane. This is discussed in detail in \citeA{PUL91}. \citeA{KAL15} claims that the events can still be compared, even though the spatial scales are underestimated. Using the T96 model, \citeA{KAL15} determined that the arcs map to the equatorial plane mostly in the range of 9-12 $R_E$, with field strengths less than 20 nT. The relevant optical analysis plots are shown below in Figure \ref{Figure5}.

\begin{figure}[h!]
	\centering
	\includegraphics[width=0.7\textwidth]{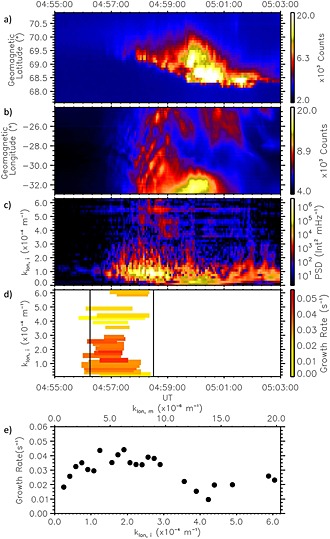}
	\caption{Optical analysis performed by \cite{KAL15} for Gillam substorm on 2 October 2011. (a.) North-south (geomagnetic) keogram to show auroral brightening and poleward propagation. (b.) East-west (geomagnetic) keogram showing time-evolution of periodic structures during substorm onset. (c.) Power spectral density as a function of $k_{y,E}$. (d.) Periods of exponential growth for each $k_{y,E}$, with growth rate denoted by color. Substorm onset interval is given by the vertical lines. Only wavenumbers that grow for over 30 s and start within one standard deviation of the median start time were used in the analysis. (e.) Growth rate as a function of wavenumbers, with ionospheric wavenumbers $k_{y,E}$ below and magnetospheric wavenumbers $k_y$ above. This figure is reproduced from \citeA{KAL15}.}
	\label{Figure5}
\end{figure}

Of the two instabilities which were not ruled out by the \citeA{KAL15} analysis, the shear flow-ballooning instability provided the best explanation of the observed beading results, corroborating previous findings along these lines \cite{FRI00}. This instability was characterized in the form used by \citeA{KAL15} in \citeA{VOR97}. This instability is a hybrid of the Kelvin-Helmholtz and Rayleigh-Taylor instabilities with larger growth rates operating on shorter growth time scales than pure Kelvin-Helmholtz instability. The former are driven by shear flows and the latter by earthward pressure gradients. An extensive linear analysis of such hybrid instabilities and their relation to substorms has been conducted by \citeA{YAM08,YAM09}. In particular, it was found that the hybrid waves are driven by shear flows in the presence of an earthward particle energy density gradient. The auroral arc is tied to the boundary between the stretched field lines and the dipolarized field lines at the inner edge of the near-Earth plasma sheet. This is where pressure gradients are most relevant. The spatial scale of the shear flow-ballooning instability varies inversely as the size of the shear flow region. \citeA{KAL15} determined that for this instability, the growth rates peak at 0.2 Hz in the wavenumber regions specified above.

After setting up a simple geometric wedge model following \citeA{WOLF18} for which perturbations will entail earthward flowing waves which carry the effects of the magnetospheric disturbance back to the ionosphere, we derive a wave equation for the plasma wedge which differs from that of both his original paper, \cite{WOLF18}, and that from which \citeA{KAL15} extracted the equation governing the shear flow-ballooning instability, namely \citeA{VOR97}. \citeA{VOR97} treated the coupling of shear flow and pressure gradient instabilities, but incorrectly perturbed the momentum equation (see Appendix A). \citeA{WOLF18}, on the other hand, did not treat the shear flow effect, thus obtaining low-frequency buoyancy waves which are not coupled to shear flow (see Section 5). Either of these alterations shifts the growth rates and, more importantly, fails to capture some of the essential qualitative features of the instability mechanism.

We will begin with the linearization of the MHD field equations from which all equations under consideration can be derived by the addition of various constraints and assumptions. The continuity equation is an unnecessary constraint if one utilizes the flux tube volume given in terms of the magnetic field strength. Upon combining these equations to obtain an ordinary radial differential equation for the radial component of the velocity (Sections 3.2 - Section 4), we take several limits to obtain a reduced low-frequency shear flow-interchange wave equation (Section 5). Throughout, we pause to mention the equation obtained by \citeA{WOLF18} for the buoyancy waves in the absence of velocity shear, in general and under the same limits. Appendix A contains a discussion of the way in which \citeA{VOR97} obtained the shear flow-ballooning wave equation by a particular misuse of the momentum equation (Section 3.1) in his linearization procedure. 

The primary result is the shear flow-interchange wave equation and what it entails.  The shear flow-interchange instability is the relevant instability which occurs in the magnetotail prior to substorm onset. The shear flow-ballooning instability equation is incorrect (resulting from a combination of inconsistencies and inappropriate assumptions in the perturbation of the momentum equation), and therefore eliminated as a proposal, and the interchange wave equation for buoyancy waves absent shear flow lacks the generality of the full shear flow-interchange wave equation. Shear flow-interchange instability should replace the \citeA{VOR97} shear flow-ballooning instability proposals to explain magnetospheric phenomena in the appropriate limits. More is said about this in Section 5, as it is a somewhat subtle issue. What was a destabilizing ballooning term is really seen to be replaced by a stabilizing interchange term in the region of interest. The shear flow couples to the interchange instability in a way which reduces the growth rates relative to shear flow-ballooning instability, and a fortiori to pure shear flow (Kelvin-Helmholtz) instability. Local WKB stability analysis is performed in the regions under consideration, and spatial scale of the instability is determined.

In summary, it appears that a shear flow-interchange instability in the midnight region of the nightside magnetopause is the most plausible link in the causal chain of events which initiate substorm onset via earthward traveling shear flow-buoyancy waves, and results in structures in the aurorae in the E-layer of the ionosphere. After perturbations drive an instability, the linear equations and dispersion relations become invalid. Nonlinear analysis will be developed in future work on the instability.

\section{Wedge Model for Local Nightside Geomagnetic Tail Plasma}

\begin{figure}[h!]
	\centering
	\textbf{a)} \\
	\includegraphics[width=0.4\textwidth]{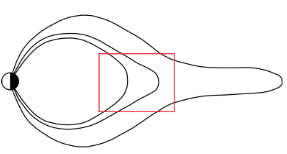} \\
	\textbf{b)} \\
	\includegraphics[width=0.4\textwidth]{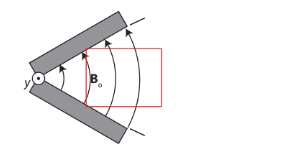}
	\caption{Near-earth nightside magnetosphere and wedge model. (a.) Nightside region of magnetosphere showing equilibrium magnetic field lines shortly prior to substorm onset. (b.) Diagram of the coordinate system for the wedge-shaped region of the magnetosphere under consideration. Field lines are approximated as concentric arcs. The $y$-axis is that of standard SM or GSM coordinates, perpendicular to the magnetic dipole and the earth-sun line. Distance from the $y$-axis is given by the $r$-coordinate, which specifies the distance of the tubes of magnetic flux from the center of the magnetosphere. Red boxes surround coincident regions in nightside magnetosphere and wedge model. Part (b.) is reproduced from \cite{WOLF18}.}
	\label{Figure6}
\end{figure}

First, we set up a cylindrical coordinate system in the near-earth nightside plasma sheet, seen to the right in Figure \ref{Figure6}. The center of the magnetosphere is taken to be the origin of cylindrical coordinates $(r,\phi ,y)$. The $y$-axis is that of standard SM or GSM coordinates, perpendicular to the magnetic dipole and the earth-sun line. Distance from the $y$-axis is given by the $r$-coordinate, which specifies the distance of the tubes of magnetic flux from the center of the magnetosphere, and hence the local magnetic curvature. The transformation to the coordinates of \citeA{VOR97} is simple in the local plasma sheet region, but globally aligns more naturally with the magnetospheric structures of interest.

Following \citeA{WOLF18}, the simplest geometry has been chosen which still allows magnetic tension to support magnetic buoyancy oscillations to drive earthward flow, so that analytical solution for the eigenvalues is possible. Our more general equation includes coupling to shear flow velocity without sacrifice to this point. The two places at the upper and lower $\phi$-boundary (at $\pm \Delta \phi /2$) of the wedge represent the northern and southern ionospheres, taken by approximation to have no conductance owing to the absence of a field-aligned current in the model. The results of this oversimplification on thin filament oscillations for buoyancy waves in the absence of velocity shear are investigated in \citeA{WOLF12-1, WOLF12-2, WOLF18}. The primary effect is to reduce the resistive damping of modes, which is not insignificant for realistic conductances on field lines in the inner magnetosphere. It is worth noting that damping becomes less significant for longer plasma sheet field lines with higher mass density owing to inertial effects.

\begin{figure}[h!]
	\centering
	\textbf{a)} \\
	\includegraphics[width=0.4\textwidth]{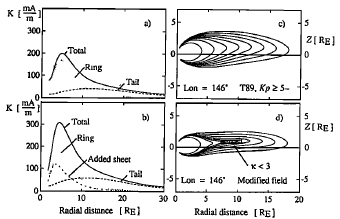} \\
	\textbf{b)} \\
	\includegraphics[width=0.4\textwidth]{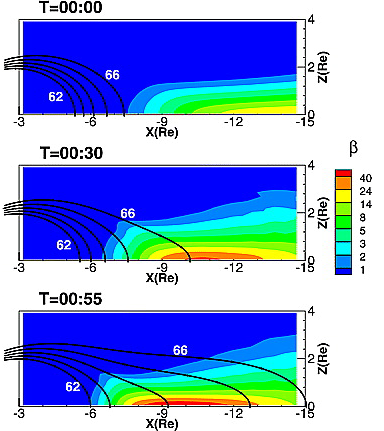}
	\caption{Plots of substorm growth phase current intensities and field line mappings. (a.) On the left, integrated current intensities (total, ring, and tail currents) over the current sheet thickness in the ISEE-1 meridian [a] in the basic T89 model, and [b] in Pulkkinen modified model with maximal parameter values. On the right, field line projections to the ISEE-1 meridian (146$^{\circ}$) computed using [c] the basic T89 model and [d] the Pulkkinen model. This figure is reproduced from \citeA{PUL91}. (b.) Plasma $\beta$ and magnetic field line mappings (every degree from 62$^{\circ}$ to 66$^{\circ}$ latitude) in the midnight meridian plane at T = 0, 30, and 55 min during the growth phase from the Rice Convection Model. This figure is reproduced from \citeA{YAN11}.}
	\label{Figure7}
\end{figure}

We will for the background parameters use a model which has more appropriately stretched field lines for the substorm growth phase than the T96 model \cite{TSY95, TSY96, TSY96-M}. Magnetic field lines get stretched substantially tailward during the growth phase prior to substorm onset \cite{PUL91, YAN11}. Plots of field line stretching during the growth phase are shown in Figure \ref{Figure7} below from three different models.

It is worth mentioning in passing that the lack of resistivity and the fixed field lines also exclude the necessary conditions for a characterization of any tearing modes which may be triggered by the shear flow-interchange instability in the region where the field lines are highly stretched. Since this instability will operate on longer timescales, it is left untreated, and as an independent hypothetical link in the sequence of events which constitute the substorm.

The system is taken to be at rest in equilibrium. Background equilibrium quantities are labeled with ``0'' subscripts, and ``$\delta$'' signifies perturbations. Magnetic field lines are approximated by concentric circles, and density, pressure, and magnetic field strength vary radially. We consider small perturbations which do not induce motion in the $\phi$-direction, so that $k_\phi = k_{\|} = 0$. Note that this implies that the field lines remain unperturbed. The pressure dynamics are modeled as adiabatic, entropy $K \coloneqq P V^\Gamma$ constant, with adiabatic gas constant $\Gamma = \left( f+2 \right)/f$, where $f$ is the number of degrees of freedom, and $V(r)=r \Delta \phi /B(r)$ for the flux tube volume. Equilibrium force balance is given by:
\bq
	- \frac{\p}{\p r} \left( P_0 + \frac{B_0^2}{2\mu_0} \right) + \frac{B_0^2}{\mu_0 r} = 0.
\eq
Note that the flux tube has a curvature towards the center of the earth in the equatorial region, and the flux tube radius $r$ is just the local radius of magnetic curvature.

Background parameters take the following form in this model:
\bqy
	\vec{B}_0 &=& B_0(r) \, \hat{\phi} \\
	\vec{v}_0 &=& v_0(r) \, \hat{y} \\
	\rho_0 &=& \rho_0(r) \\
	P_0 &=& P_0(r).
\eqy
So we assume that all equilibrium quantities are static and depend only on radius. The velocity has the form of an axially-directed $\vec{E} \times \vec{B}$ shear flow. More details about this model, such as the specific radial profiles of the background parameters and some of the perturbations which result under the assumption of no shear velocity (only buoyancy waves) can be found in \citeA{WOLF18}.

\section{Linear Dynamics of Geomagnetic Tail Plasma}

We begin with the MHD field equations:
\bqy
	\frac{\p}{\p t} \left( \frac{P}{\rho^\Gamma} \right) + \vec{v} \cdot \nabla \left( \frac{P}{\rho^\Gamma} \right) &=& 0 \\
	\frac{\p \vec{B}}{\p t} - \nabla \times \left( \vec{v} \times \vec{B} \right) &=& 0 \\
	\rho \frac{\p \vec{v}}{\p t} + \rho \vec{v} \cdot \nabla \vec{v} &=& - \nabla \left( P + \frac{B^2}{2 \mu _0} \right) + \frac{\vec{B} \cdot \nabla \vec{B}}{\mu_0} \label{momentum} \\
	\frac{\p \rho}{\p t} + \nabla \cdot \left( \rho \vec{v} \right) &=& 0.
\eqy

\begin{figure}[h!]
	\centering
	\includegraphics[width=0.7\textwidth]{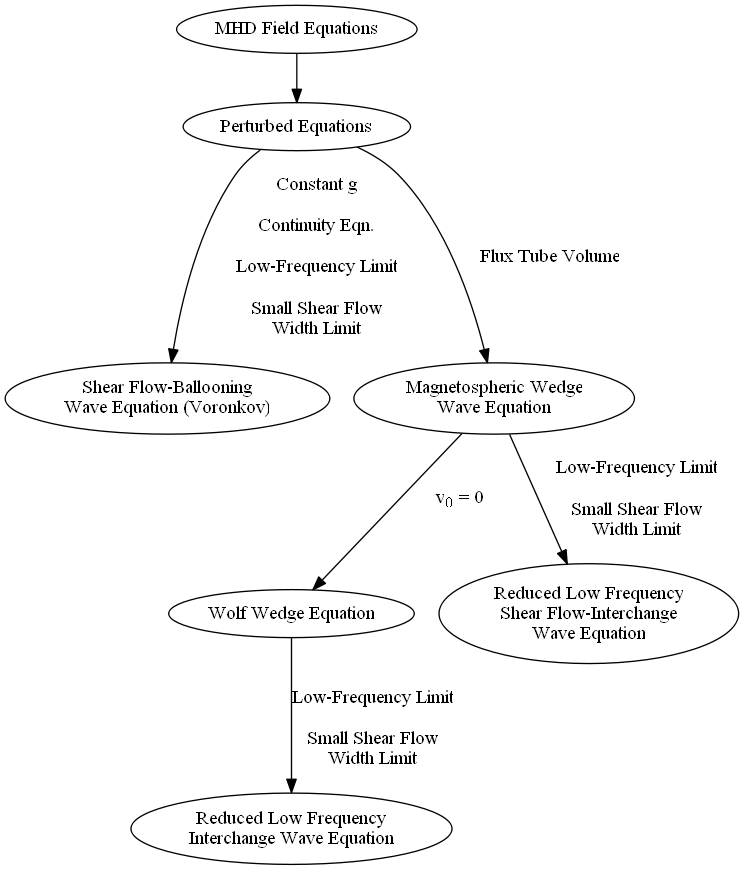}
	\caption{Inferential Pathway Flowchart. Nodes signify important equations or systems of equations and arrows signify physical assumptions made in the transition between nodes. The assumptions for a particular transition are specified to the right of the given arrow. The equation which most plausibly supplies an explanation for the instabilities which occur in the magnetotail before substorm onset is the reduced low-frequency shear flow-interchange wave equation.}
	\label{Figure8}
\end{figure}

A flowchart \ref{Figure8} of the inferential pathways and corresponding assumptions for all equations to be analyzed is included below for reference. It should facilitate a global view of the interrelations.

\subsection{Linearized MHD Equations for Wedge Flux Tube}

Now, we apply the system of equations to the wedge formalism. Let's survey each equation and discuss.

We can from the start bypass the continuity equation which \citeA{VOR97} uses as an additional constraint by assuming the form of the density to be that of a flux tube:
\bq
	\rho = \frac{B}{r \Delta \phi}
\eq
in the adiabatic pressure dynamics equation. Note that the $\Delta \phi$ is a constant. It is fitting to define the entropy:
\bq
	K := P V^\Gamma = \frac{P}{\rho^\Gamma}
\eq
for future use in eliminating pressure and more intuitively representing interchange dynamics of the flux tubes under consideration. The adiabatic pressure dynamics using the convective time-derivative:
\bq
	\frac{D}{Dt} \left( \frac{P}{\rho^\Gamma} \right) = 0,
\eq
thus take the following form for a flux tube in the wedge formalism:
\bq
	\frac{D}{Dt} \left( P \left( \frac{r}{B} \right)^\Gamma \right) = 0.
\eq
Substituting the total fields and linearizing, we obtain the equation which governs the pressure fluctuation dynamics:
\bq
	\left( \frac{\p}{\p t} + v_0 \frac{\p}{\p y} \right) \left( \frac{\delta P}{P_0} - \Gamma \frac{\delta B}{B_0} \right) + \frac{K_0'}{K_0} \delta v_r = 0. \label{1}
\eq
Henceforth, partial radial derivatives will be indicated by primes when convenient.

The magnetic field dynamics in the flux tube are governed by Faraday's law with $\vec{E} = - \vec{v} \times \vec{B}$, owing to the high conductivity in the region:
\bq
	\frac{\p \vec{B}}{\p t} = \nabla \times \left( \vec{v} \times \vec{B} \right).
\eq
Substituting the total fields and linearizing, we obtain:
\bqy
	\frac{\p \delta B_r}{\p t} &=& - v_0 \frac{\p \delta B_r}{\p y} \label{2} \\
	\frac{\p \delta B_\phi}{\p t} &=& - v_0 \frac{\p \delta B_\phi}{\p y} - B_0 \frac{\p \delta v_y}{\p y} - B_0 \delta v_r' - B_0' \delta v_r \label{3} \\
	\frac{\p \delta B_y}{\p t} &=& \frac{v_0}{r} \delta B_r + v_0' \delta B_r + v_0 \delta B_r'. \label{4}
\eqy

Now, we examine the momentum equation, which governs the plasma dynamics:
\bq
	\rho \frac{\p \vec{v}}{\p t} + \rho \vec{v} \cdot \nabla \vec{v} = - \nabla \left( P + \frac{B^2}{2\mu_0} \right) + \frac{\vec{B} \cdot \nabla  \vec{B}}{\mu _0}.
\eq
Substituting the total fields and linearizing, we obtain:
\bqy
	\rho_0 \frac{\p \delta v_r}{\p t} + \rho_0 v_0 \frac{\p \delta v_r}{\p y} &=& - \frac{\p }{\p r} \left( \delta P + \frac{B_0}{\mu_0} \delta B_\phi \right) - \frac{2 B_0}{\mu_0 r} \delta B_\phi \label {5} \\
	\rho_0 \frac{\p \delta v_\phi}{\p t} + \rho_0 v_0 \frac{\p \delta v_\phi}{\p y} &=& \frac{B_0}{\mu_0 r} \delta B_r + \frac{B_0'}{\mu _0} \delta B_r \label{6} \\
	\rho_0 \frac{\p \delta v_y}{\p t} + \rho_0 v_0 \frac{\p \delta v_y}{\p y} + \rho_0 v_0' \delta v_r &=& - \frac{\p \delta P}{\p y} - \frac{B_0}{\mu_0} \frac{\p \delta B_\phi}{\p y}, \label{7}
\eqy
which governs the plasma acceleration.

In the wedge formalism, the continuity equation is unnecessary, as the density is already expressed in terms of the magnetic field. Thus, the density perturbations are implicit in magnetic field fluctuations of flux tubes via flux freezing.

\section{Magnetospheric Wave Equation for Plasma Wedge}

Now, we assume all perturbations take the form of axially propagating waves $e^{i k_y y - i \omega t}$ in the plasma sheet, denoting the Doppler-shifted frequency:
\bq
	\widetilde{\omega}(r) \coloneqq \omega - k_y v_0(r),
\eq
as these are the waves which will map back to the ionosphere to cause the longitudinally-directed auroral beads. Note that this converts primes into total rather than partial radial derivatives.

Substituting this form into our self-consistent set of dynamical equations (\ref{1})-(\ref{7}), we obtain:
\bqy
	i \widetilde{\omega} \left( \frac{\delta P}{P_0} - \Gamma \frac{\delta B}{B_0} \right) &=& \frac{K_0'}{K_0} \delta v_r \label{A} \\
	- i \widetilde{\omega} \delta B_r &=& 0 \label{B} \\
	- i \widetilde{\omega} \delta B_\phi &=& -i k_y B_0 \delta v_y - B_0' \delta v_r - B_0 \delta v_r' \label{C} \\
	-i \omega \delta B_y &=& \frac{v_0}{r} \delta B_r + v_0' \delta B_r + v_0 \delta B_r' \label{D} \\
	- i \widetilde{\omega} \rho_0 \delta v_r &=& - \frac{\p}{\p r} \left( \delta P + \frac{B_0}{\mu_0} \delta B_\phi \right) - \frac{2 B_0}{\mu_0 r} \delta B_\phi \label{E} \\
	-i \widetilde{\omega} \rho_0 \delta v_\phi &=& \frac{B_0}{\mu_0 r} \delta B_r + \frac{B_0'}{\mu_0} \delta B_r \label{F} \\
	-i \widetilde{\omega} \rho_0 \delta v_y + \rho_0 v_0' \delta v_r &=& - i k_y \delta P-i k_y \frac{B_0}{\mu_0} \delta B_\phi \label{G}.
\eqy
We can now see that (\ref{B}), (\ref{D}), and (\ref{F}) imply the following perturbation components:
\bqy
	\delta \vec{B} &=& \delta B_\phi \hat{\phi} \\
	\delta \vec{v} &=& \delta v_r \hat{r} + \delta v_y \hat{y}.
\eqy
So three of the equations are now implicitly taken into account, and from the remaining of Eqs. (\ref{A})-(\ref{G}) we obtain the following system:
\bqy
	i \widetilde{\omega} \left( \frac{\delta P}{P_0} - \Gamma \frac{\delta B}{B_0} \right) &=& \frac{K_0'}{K_0} \delta v_r \\
	i \widetilde{\omega} \frac{\delta B_\phi}{B_0} &=& i k_y \delta v_y + \delta v_r' + \frac{B_0'}{B_0} \delta v_r \\
	i \widetilde{\omega} \rho_0 \delta v_r &=& \frac{\p}{\p r} \left( \delta P + \frac{B_0}{\mu_0} \delta B_\phi \right) + \frac{2 B_0}{\mu_0 r} \delta B_\phi \\
	i \widetilde{\omega} \rho_0 \delta v_y - \rho_0 v_0' \delta v_r &=& i k_y \delta P + i k_y \frac{B_0}{\mu_0} \delta B_\phi.
\eqy

From now on, it will be convenient to make frequent use of the Alfv{\'e}n speed, sound speed, and fast mode wave speeds given by $c_A^2 \coloneqq B_0^2/ \mu_0 \rho_0$, $c_s^2 \coloneqq \Gamma P_0 / \rho_0$, and $c_f^2 \coloneqq c_A^2 + c_s^2$, respectively.

Now, for convenience, the equilibrium force balance equation can be recast as a condition to eliminate $B_0$ in lieu of $K_0$:
\bq
	\frac{B_0'}{B_0} = \frac{1}{c_f^2} \left( - \frac{c_s^2}{\Gamma} \frac{K_0'}{K_0} + \frac{c_s^2 - c_A^2}{r} \right),
\eq
which will make manifest the interchange instability and its stability conditions.

Eliminating $\delta v_y$, $\delta P$, and $\delta B_\phi$, we obtain the differential equation for the radial velocity fluctuations $\delta v_r$, written in a form which most resembles that of \citeA{WOLF18}:
\bqy
	\widetilde{\omega}^2 \delta v_r &=& \frac{\widetilde{\omega}}{\rho_0} \frac{d}{dr} \left( - \frac{\delta v_r \rho_0 \widetilde{\omega} \left( c_s^2 - c_A^2 \right)}{r \left( \widetilde{\omega}^2 - k_y^2 c_f^2 \right)} - \frac{d \delta v_r}{dr} \frac{c_f^2 \rho_0 \widetilde{\omega}}{\widetilde{\omega}^2 - k_y^2 c_f^2} - \frac{\delta v_r \rho_0 v_0' k_y c_f^2}{\widetilde{\omega}^2 - k_y^2 c_f^2} \right) \nonumber \\
	& & + \frac{2 c_A^2}{r} \left( \frac{\delta v_r}{\Gamma c_f^2} \left[ c_s^2 \frac{K_0'}{K_0} - \frac{\widetilde{\omega}^2 \Gamma \left( c_s^2 - c_A^2 \right)}{r \left( \widetilde{\omega}^2 - k_y^2 c_f^2 \right)} \right] - \frac{d \delta v_r}{dr} \frac{\widetilde{\omega}^2}{\widetilde{\omega}^2 - k_y^2 c_f^2} - \frac{\delta v_r k_y \widetilde{\omega} v_0'}{\widetilde{\omega}^2 - k_y^2 c_f^2} \right).
\eqy
Indeed, in this form, it is easy to see that dropping velocity shear terms yields precisely the equation in \citeA{WOLF18}:
\bqy
	\omega^2 \delta v_r &=& \frac{\omega}{\rho_0} \frac{d}{dr} \left( - \frac{\delta v_r \rho_0 \omega \left( c_s^2 - c_A^2 \right)}{r \left( \omega^2 - k_y^2 c_f^2 \right)} - \frac{d \delta v_r}{dr} \frac{c_f^2 \rho_0 \omega}{\omega^2 - k_y^2 c_f^2} \right) \nonumber \\
	& & + \frac{2 c_A^2}{r} \left( \frac{\delta v_r}{\Gamma c_f^2} \left[ c_s^2 \frac{K_0'}{K_0} - \frac{\omega^2 \Gamma \left( c_s^2 - c_A^2 \right)}{r \left( \omega^2 - k_y^2 c_f^2 \right)} \right] - \frac{d \delta v_r}{dr} \frac{\omega^2}{\omega^2 - k_y^2 c_f^2} \right).
\eqy
Note that the frequencies are no longer Doppler-shifted (there is no shear velocity to supply the shift!). The objective of \citeA{WOLF18} was to study buoyancy waves in the magnetosphere, and velocity shear terms were thus neglected in order to facilitate a clearer understanding of the interchange-induced buoyancy waves, with buoyancy force arising from magnetic tension rather than gravity. This equation still describes both fast mode longitudinal and buoyancy waves in the plasma wedge, but the former are easily eliminated, which we will demonstrate in what follows.

With all derivatives performed, and all terms expanded, the magnetospheric wave equation takes the following form:
\bq
	\delta v_r'' + B(r) \delta v_r' + C(r) \delta v_r = 0,
\eq
with coefficients:
\bqy
	B(r) &\coloneqq& \frac{1}{r} + \frac{\widetilde{\omega}^2}{\widetilde{\omega}^2 - k_y^2 c_f^2} \frac{c_s^2}{c_f^2} \frac{P_0'}{P_0} + \frac{\widetilde{\omega}^2}{\widetilde{\omega}^2 - k_y^2 c_f^2} \frac{2 c_A^2}{c_f^2} \frac{B_0'}{B_0} - \frac{k_y^2 c_f^2}{\widetilde{\omega}^2 - k_y^2 c_f^2} \frac{\rho_0'}{\rho_0} + 2 \frac{k_y v_0'}{\widetilde{\omega}} \frac{\widetilde{\omega}^2}{\widetilde{\omega}^2 - k_y^2 c_f^2} \\
	C(r) &\coloneqq& \frac{k_y v_0''}{\widetilde{\omega}} + 2 \frac{k_y v_0'}{\widetilde{\omega}} \frac{k_y v_0'}{\widetilde{\omega}} \frac{\widetilde{\omega}^2}{\widetilde{\omega}^2 - k_y^2 c_f^2} - \frac{k_y v_0'}{\widetilde{\omega}} \frac{k_y^2 c_f^2}{\widetilde{\omega}^2 - k_y^2 c_f^2} \frac{\rho_0'}{\rho_0} + \frac{k_y v_0'}{\widetilde{\omega}} \frac{\widetilde{\omega}^2}{\widetilde{\omega}^2 - k_y^2 c_f^2} \frac{c_s^2}{c_f^2} \frac{P_0'}{P_0} \nonumber \\
	& & + \frac{k_y v_0'}{\widetilde{\omega}} \frac{\widetilde{\omega}^2}{\widetilde{\omega}^2 - k_y^2 c_f^2} \frac{2 c_A^2}{c_f^2} \frac{B_0'}{B_0} + \frac{k_y v_0'}{\widetilde{\omega}} \left( 1 + 2 \frac{c_s^2 - c_A^2}{c_f^2} \frac{k_y^2 c_f^2}{\widetilde{\omega}^2 - k_y^2 c_f^2} \right) \frac{1}{r} - \frac{(c_s^2-c_A^2)^2}{c_f^4} \frac{1}{r^2} \nonumber \\
	& & - \frac{k_y^2 c_f^2}{\widetilde{\omega}^2 - k_y^2 c_f^2} \frac{c_s^4 - c_A^4}{c_f^4} \frac{1}{r} \frac{\rho_0'}{\rho_0} + \frac{\widetilde{\omega}^2 - 2 k_y^2 c_A^2}{\widetilde{\omega}^2 - k_y^2 c_f^2} \frac{c_s^2}{c_f^2} \frac{1}{r} \frac{P_0'}{P_0} - \frac{\widetilde{\omega}^2 - 2 k_y^2 c_s^2}{\widetilde{\omega}^2 - k_y^2 c_f^2} \frac{2 c_A^2}{c_f^2} \frac{1}{r} \frac{B_0'}{B_0} \nonumber \\
	& & - \frac{\widetilde{\omega}^2 - k_y^2 c_f^2}{\widetilde{\omega}^2} \frac{2}{\Gamma r} \frac{c_A^2 c_s^2}{c_f^4} \frac{K_0'}{K_0} + \frac{\widetilde{\omega}^2 - k_y^2 c_f^2}{c_f^2}.
\eqy

\section{Reduced Low Frequency Wedge Wave Equation}

We now eliminate fast modes, assuming $\widetilde{\omega}^2 \ll k_y^2 c_f^2$, retaining only those modes which play a substantial role in substorm onset. Upon doing so, we obtain:
\bq
	\delta v_r'' + \left( \frac{1}{r} + \frac{\rho_0'}{\rho_0} - \frac{2 k_y v_0'}{\widetilde{\omega}} \frac{\widetilde{\omega}^2}{k_y^2 c_f^2} \right) \delta v_r' + C_{lf}(r) \delta v_r = 0
\eq
with coefficient:
\bqy
	C_{lf}(r) &\coloneqq& \frac{k_y v_0''}{\widetilde{\omega}} - 2 {\left( \frac{v_0'}{c_f} \right)}^2 + \frac{k_y v_0'}{\widetilde{\omega}} \left( 1 - 2 \frac{c_s^2 - c_A^2}{c_f^2} \right) \frac{\rho_0'}{\rho_0} - \frac{k_y v_0'}{\widetilde{\omega}} \frac{1}{r} - \frac{(c_s^2 - c_A^2)^2}{c_f^4} \frac{1}{r^2} \nonumber \\
	& & + \frac{c_s^4 - c_A^4}{c_f^4} \frac{1}{r} \frac{\rho_0'}{\rho_0} + \frac{2 c_s^2 c_A^2}{c_f^4} \frac{1}{r} \frac{P_0'}{P_0} - \frac{4 c_s^2 c_A^2}{c_f^4} \frac{1}{r} \frac{B_0'}{B_0} + \frac{k_y^2 c_f^2}{\widetilde{\omega}^2} \frac{2}{\Gamma r} \frac{c_A^2 c_s^2}{c_f^4} \frac{K_0'}{K_0} - k_y^2.
\eqy
We also assume that $\delta v_r$ vary on length scales $\sim k_r^{-1}$, which is small compared to $r$ and variations in $B_0 \sim L$ (from which the scale for $P_0$ follows from equilibrium force balance). Recall that this effect is even more pronounced due to the field line stretching which occurs during the substorm growth phase prior to the instability. In short, we assume small shear flow width $\delta =\sim 1/k_r \ll L$. Care must be taken to ensure that terms which involve ratios of the small parameters are not hastily dropped. The density gradient terms should not be dropped, as it is quite possible that there is a sharp plasmapause prior to substorm onset.

The resulting reduced low frequency equation gives the long wavelength shear flow-buoyancy waves in the nightside wedge:
\bq
	\delta v_r'' + \left( \frac{\rho_0'}{\rho_0} - \frac{2 k_y v_0'}{\widetilde{\omega}} \frac{\widetilde{\omega}^2}{k_y^2 c_f^2} \right) \delta v_r' + \left( \frac{k_y v_0''}{\widetilde{\omega}} + \frac{k_y v_0'}{\widetilde{\omega}} \frac{\rho_0'}{\rho_0} - 2 {\left( \frac{v_0'}{c_f} \right)}^2 + \frac{k_y^2 c_f^2}{\widetilde{\omega}^2} \frac{2}{\Gamma r} \frac{c_A^2 c_s^2}{c_f^4} \frac{K_0'}{K_0} - k_y^2 \right) \delta v_r = 0 \label{SFI},
\eq
with the $\vec{E} \times \vec{B}$ shear flow velocity $v_0(r)$ and local magnetic curvature determining the dynamic stability conditions. This equation for low frequency waves in the wedge captures the most general dynamical phenomena relevant to the causal chain of events which we aim to describe. The first term in the first derivative coefficient is an inertial damping term. The first term in the zeroth derivative coefficient is the shear flow instability term, and the fourth term is the interchange instability term that supplies the buoyancy frequency.

Though it does not pertain to the more general analysis at hand, it should be mentioned that these limits, taken in the appropriate order, agree with those in \citeA{WOLF18}, barring what appear to be minor typographical errors (as seen by a unit check) on his part (equations (19) and (20) of \citeA{WOLF18}). The buoyancy frequency, which was thoroughly discussed in \citeA{WOLF18}, is given by the next-to-last term in our equation. Let us perform this check. Dropping the shear velocity terms, we obtain:
\bq
	\delta v_r'' + \frac{\rho_0'}{\rho_0} \delta v_r' + \left( \frac{1}{\omega^2} \frac{2}{\Gamma r} \frac{c_A^2 c_s^2}{c_f^2} \frac{K_0'}{K_0} - 1 \right) k_y^2 \delta v_r = 0
\eq
Thus the first term in parentheses yields immediately the buoyancy frequency for waves in a wedge:
\bq
	\omega_b^2(r) = \frac{2}{\Gamma r} \frac{c_A^2 c_s^2}{c_f^2} \frac{K_0'}{K_0}
\eq
It is demonstrated in \citeA{WOLF18} that this is just the oscillation frequency of thin magnetic filaments in the wedge. The speed $c_A c_s/c_f$ is just that of the slow mode buoyancy waves which result from pure interchange oscillations. Notice also that the frequencies are no longer Doppler-shifted, as there is no fluid velocity to supply the shift.

Recast in the above notation, the \citeA{VOR97} result (obtained in a methodologically similar way corresponding (roughly) to limits taken in Section 5, though with the additional assumption that density gradient scales are large) utilized by \citeA{KAL15} is:
\bq
	\delta v_r'' + \left( \frac{v_0''}{k_y \widetilde{\omega}} - \frac{g}{\widetilde{\omega}^2} \frac{\rho_0'}{\rho_0} - \frac{1}{\widetilde{\omega}^2} \frac{g^2}{c_f^2} - 1 \right) k_y^2 \delta v_r = 0.
\eq
As written by \citeA{VOR97}, this has the form:
\bq
	\delta v_r'' + \left( \frac{v_0''}{k_y \widetilde{\omega}} + \frac{W}{\widetilde{\omega}^2} - 1 \right) k_y^2 \delta v_r = 0.
\eq
, where:
\bq
	W := - \frac{g \rho_0'}{\rho_0} - \frac{g^2}{c_f^2},
\eq
with effective acceleration defined above. The term W obtained by \citeA{VOR97} was taken to be an analog of the buoyancy frequency, $\omega_b^2$. Near the inner edge of the plasma sheet, at approximately 9-12 $R_E$, the analysis of \citeA{KAL15} indicates that these terms are destabilizing, whereas the above analysis reveals these terms to be stabilizing. This is due to differences between ballooning and interchange, where the former are often treated as localized and the latter is globally distributed along the magnetic field lines. The interchange instability results, rather than ballooning, since the unstable mode does not perturb the background equilibrium magnetic field.

These discussions are reproduced in the respective appendices so that they appear both for comparison with the reduced wedge wave equation and as part of a complete discussion of each instability.

\section{Examination of Stability Conditions}

Stability analysis for the shear flow-interchange modes resulting from equation (\ref{SFI}) is performed locally using WKB approximation. Since the length scales for several background gradients in the region have been assumed small in comparison to the shear flow width, we can examine the propagation and growth of a wave packet in the region centered on some wavenumber, with a width much smaller than the gradient length scales. Assuming that the growth rate is smaller than the frequency, the wave packet can propagate for some time governed by the linearized equations. Upon undergoing about 10 e-foldings, the linearized treatment must be replaced by the full nonlinear analysis. The sub-case interchange instability (treated in more detail in \citeA{WOLF18}) is discussed in Appendix B for comparison. While we do not discuss stability conditions for the \citeA{VOR97}, it is worth noting that what were ballooning terms under that treatment are seen to be interchange terms. The phenomenological import of this is that the former are localized, whereas the latter are globally distributed along the magnetic field lines.

\subsection{Shear Flow-Interchange Instability}

In preparation for the obtainment of a dispersion relation, let us recast the differential equation in a simpler form to see the wavelike properties more clearly by extracting the radius-dependent prefactor as follows. For convenience, the equation for shear flow-interchange modes again is:
\bq
	\delta v_r'' + \left( \frac{\rho_0'}{\rho_0} - \frac{2 k_y v_0'}{\widetilde{\omega}} \frac{\widetilde{\omega}^2}{k_y^2 c_f^2} \right) \delta v_r' + \left( \frac{k_y v_0''}{\widetilde{\omega}} + \frac{k_y v_0'}{\widetilde{\omega}} \frac{\rho_0'}{\rho_0} - 2 {\left( \frac{v_0'}{c_f} \right)}^2 + \frac{k_y^2 c_f^2}{\widetilde{\omega}^2} \frac{2}{\Gamma r} \frac{c_A^2 c_s^2}{c_f^4} \frac{K_0'}{K_0} - k_y^2 \right) \delta v_r = 0.
\eq
We perform this formal simplification as follows. Let:
\bqy
 	\delta v_r &\coloneqq& \alpha(r) \delta u_r, \\
 	\mbox{so that:} \hspace{3em} & & \nonumber \\
 	\delta v_r' &=& \alpha' \delta u_r + \alpha \delta u_r' \\
 	\delta v_r'' &=& \alpha'' \delta u_r + 2 \alpha' \delta u_r' + \alpha \delta u_r''.
\eqy
Now, we choose an $\alpha$ such that the prefactor to $\delta u_r'$ is zero in the new variables:
\bq
	\alpha(r) = \frac{Const.}{\sqrt{\rho_0}} exp \left( \int{\frac{\widetilde{\omega} v_0'}{k_y c_f^2}} \right),
\eq
whereby:
\bqy
	\frac{\alpha'}{\alpha} &=& \frac{k_y v_0'}{\widetilde{\omega}} \frac{\widetilde{\omega}^2}{k_y^2 c_f^2} - \frac{1}{2} \frac{\rho_0'}{\rho_0} \\
	\frac{\alpha''}{\alpha} &=& \left( \frac{\alpha'}{\alpha} \right)'+ \left( \frac{\alpha'}{\alpha} \right)^2,
\eqy
so we obtain the following form:
\bq
	\delta u_r'' + \left( \frac{1}{4} {\left( \frac{\rho_0'}{\rho_0} \right)}^2 - \frac{1}{2} \frac{\rho_0''}{\rho_0} + \frac{k_y v_0''}{\widetilde{\omega}} - \frac{k_y v_0'}{\widetilde{\omega}} \frac{\rho_0'}{\rho_0} - 3 {\left( \frac{v_0'}{c_f} \right)}^2 + \frac{k_y^2 c_f^2}{\widetilde{\omega}^2} \frac{2}{\Gamma r} \frac{c_A^2 c_s^2}{c_f^4} \frac{K_0'}{K_0} - k_y^2 \right) \delta u_r = 0,
\eq
which manifestly has wave solutions (insofar as WKB approximation validates assuming the coefficients can be treated as roughly constant over the width of the wave packet):
\bq
	\delta u_r(r) \sim e^{i k_r r}.
\eq

Recall that it is the power spectral density which is obtained by Fourier transforming the ASI intensity keograms to obtain the instability growth rates. But the power spectral density is just the energy density being transported at the fluid velocity. It is the fluid velocity which will contain the growth rate, and radial and axial velocities have the same growth rate. This is just a reminder that we can straightforwardly proceed by identifying the radial fluid velocity growth rate obtained here with the empirically obtained power spectral density growth rates.

Upon converting our magnetospheric wedge wave equation into a dispersion relation for shear flow-interchange modes, with radial oscillations having wavenumber $k_r \sim 1/\delta$, keeping in mind that the radius needs Doppler-shifted, we obtain:
\bqy
	& & \left( \frac{1}{2} \frac{\rho_0''}{\rho_0} - \frac{1}{4} \left( \frac{\rho_0'}{\rho_0} \right)^2 + 3 \left( \frac{v_0'}{c_f} \right)^2 + \left( k_y^2 + k_r^2 \right) \right) \omega^2 \nonumber \\
	& & + \left( - \frac{\rho_0''}{\rho_0} + \frac{1}{2} \left( \frac{\rho_0'}{\rho_0} \right)^2 - \frac{v_0''}{v_0} - 6 \left( \frac{v_0'}{c_f} \right)^2 + \frac{v_0'}{v_0} \frac{\rho_0'}{\rho_0} - 2 \left( k_y^2 + k_r^2 \right) \right) k_y v_0 \omega \nonumber \\
	& & + \left( \frac{1}{2} \frac{\rho_0''}{\rho_0} - \frac{1}{4} \left( \frac{\rho_0'}{\rho_0} \right)^2 + \frac{v_0''}{v_0} + 3 \left( \frac{v_0'}{c_f} \right)^2 - \frac{v_0'}{v_0} \frac{\rho_0'}{\rho_0} - \frac{2}{\Gamma r} \frac{c_A^2 c_s^2}{c_f^2 v_0^2} \frac{K_0'}{K_0} + \left( k_y^2 + k_r^2 \right) \right) k_y^2 v_0^2 = 0.
\eqy
Letting $\omega = \omega_r + i \gamma$, and solving for both the frequency and the growth rate, we obtain the following:
\bqy
	\omega_r(k_r,k_y;r) &=& \frac{\frac{\rho_0''}{\rho_0} - \frac{1}{2} \left( \frac{\rho_0'}{\rho_0} \right)^2 + \frac{v_0''}{v_0} + 6 \left( \frac{v_0'}{c_f} \right)^2 - \frac{v_0'}{v_0} \frac{\rho_0'}{\rho_0} + 2 (k_y^2 + k_r^2)}{\frac{\rho_0''}{\rho_0} - \frac{1}{2} \left( \frac{\rho_0'}{\rho_0} \right)^2 + 6 \left( \frac{v_0'}{c_f} \right)^2 + 2 (k_y^2 + k_r^2)} k_y v_0 \\
	\gamma(k_r,k_y;r) &=& \pm \frac{{\left( - \left( \frac{v_0''}{v_0} - \frac{v_0'}{v_0} \frac{\rho_0'}{\rho_0} \right)^2 - \left( \frac{\rho_0''}{\rho_0} - \frac{1}{2} \left( \frac{\rho_0'}{\rho_0} \right)^2 + 6 {\left( \frac{v_0'}{c_f} \right)}^2 + 2 \left( k_y^2 + k_r^2 \right) \right) \frac{4}{\Gamma r} \frac{c_A^2 c_s^2}{c_f^2 v_0^2} \frac{K_0'}{K_0} \right)}^{1/2}}{\frac{\rho_0''}{\rho_0} - \frac{1}{2} \left(\frac{\rho_0'}{\rho_0} \right)^2 + 6 \left( \frac{v_0'}{c_f} \right)^2 + 2 (k_y^2 + k_r^2)} k_y v_0.
\eqy
Now, where growth and oscillations occur simultaneously, we should examine the situation where the growth rate is much smaller than the frequency, so that we can consider a propagating wave packet which exponentially grows as it travels until it reaches the point where nonlinearities must be considered. If the growth rate is much larger than the frequency, the instability will grow too quickly for the packet to propagate before nonlinear analysis becomes necessary. Marginal stability analysis can now be used to determine the radius and radial wavenumber for the most unstable modes (by finding growth rate extrema). The wave packet will initially be centered at the radial wavenumber and begin its propagation at the radial extremum. The axial wavenumber remains fixed during propagation, since the frequency has no y-dependence. The instability extrema will occur near radial wavenumbers:
\bq
	k_r^*(r) = \pm \sqrt{\frac{1}{4} \left( \frac{\rho_0'}{\rho_0} \right)^2 - \frac{1}{2} \frac{\rho_0''}{\rho_0} - 3 \left( \frac{v_0'}{c_f} \right)^2 - \frac{1}{2} \frac{\left( \frac{v_0''}{v_0} - \frac{v_0'}{v_0} \frac{\rho_0'}{\rho_0} \right)^2}{\frac{2}{\Gamma r} \frac{c_A^2 c_s^2}{c_f^2 v_0^2} \frac{K_0'}{K_0}} - k_y^2}
\eq
Without background profiles, the general expression for the radius at which growth rate is optimized is impossible to analytically express. The general constraint condition on background profiles which must hold at that radius is unilluminating. But we can examine the extremization condition on radius at the radial wavenumber extremum:
\bq
	\frac{2}{\Gamma r} \frac{c_A^2 c_s^2}{c_f^2 v_0^2} \frac{K_0'}{K_0} = \frac{V}{v_0} \left( \frac{v_0''}{v_0} - \frac{v_0'}{v_0} \frac{\rho_0'}{\rho_0} \right),
\eq
where V is a radially constant velocity. We can write this as an explicit definition of the new velocity:
\bq
	V \coloneqq \frac{\frac{2}{\Gamma r} \frac{c_A^2 c_s^2}{c_f^2 v_0^2} \frac{K_0'}{K_0}}{\frac{v_0''}{v_0} - \frac{v_0'}{v_0} \frac{\rho_0'}{\rho_0}} v_0.
\eq
This means that our peak wavenumber can be recast in the following convenient form:
\bq
	k_r^*(r^*) = \pm \sqrt{\frac{1}{4} \left( \frac{\rho_0'}{\rho_0} \right)^2 - \frac{1}{2} \frac{\rho_0''}{\rho_0} - 3 \left( \frac{v_0'}{c_f} \right)^2 - \frac{1}{2} \frac{v_0''}{V} + \frac{1}{2} \frac{v_0'}{V} \frac{\rho_0'}{\rho_0} - k_y^2}.
\eq
For the maximally sensitive mode, the growth rate and frequency are given by:
\bqy
	\omega_r &=& k_y (v_0 - V) \\
	\gamma &=& \mp k_y V.
\eqy
Notice that this implies that the Doppler-shifted real part becomes precisely:
\bq
	\widetilde{\omega}_r = - k_y V.
\eq
It is clear at this point that V is just the Doppler-shifted phase velocity of the propagating wave packet. Note in addition that the growth rate sign is always positive, and the choice of explicit sign cancels the implicit sign of $k_r$ so as to maintain independence of propagation direction. Thus extremized, our growth-to-frequency ratio is seen to be given by:
\bq
	\frac{\gamma}{\omega_r} = \pm \frac{V}{V - v_0},
\eq
which will be small when $v_0 \gg V$. It is under these circumstances that we will have a growing propagating wave packet. The group velocity and time variation of peak wavenumber are generally given by:
\bqy
	\dot{r} &=& \frac{\p \omega_r}{\p k_r} \\
	\dot{y} &=& \frac{\p \omega_r}{\p k_y} \\
	\dot{k_y} &=& - \frac{\p \omega_r}{\p y} \\
	\dot{k_r} &=& - \frac{\p \omega_r}{\p r}.
\eqy
For the maximally unstable mode, the packet will evolve according to the following:
\bqy
	\dot{r} &=& - \frac{4 k_y k_r^* V}{\frac{v_0''}{V} - \frac{v_0'}{V} \frac{\rho_0'}{\rho_0}} \\
	\dot{y} &=& v_0 - V - \frac{V^2}{v_0^2} \frac{4 k_y^2 v_0}{\frac{v_0''}{v_0} - \frac{v_0'}{v_0} \frac{\rho_0'}{\rho_0}} \\
	\dot{k_y} &=& 0 \\
	\dot{k_r} &=& - k_y v_0'.
\eqy
The growth factor can thus be determined for this mode:
\bq
	\int{\gamma dt} = \int{\frac{\gamma}{\dot{r}} dr} = \frac{1}{4} \int{\frac{\frac{v_0''}{V} - \frac{v_0'}{V} \frac{\rho_0'}{\rho_0}}{\sqrt{\frac{1}{4} \left( \frac{\rho_0'}{\rho_0} \right)^2 - \frac{1}{2} \frac{\rho_0''}{\rho_0} - 3 \left( \frac{v_0'}{c_f} \right)^2 - \frac{1}{2} \frac{v_0''}{V} + \frac{1}{2} \frac{v_0'}{V} \frac{\rho_0'}{\rho_0} - k_y^2}} dr}.
\eq
This is the real part of the most unstable mode's exponent during its linear evolution. When it becomes $\sim 10$, the linear approximations become invalid and nonlinear analysis becomes necessary.

\subsection{WKB Applicability Conditions}

Let us again examine the wavenumber for the most unstable mode to see if it satisfies the full applicability conditions for WKB:
\bq
	k_r^*(r^*) = \sqrt{\frac{1}{4} \left( \frac{\rho_0'}{\rho_0} \right)^2 - \frac{1}{2} \frac{\rho_0''}{\rho_0} - 3 \left( \frac{v_0'}{c_f} \right)^2 - \frac{1}{2} \frac{v_0''}{V} + \frac{1}{2} \frac{v_0'}{V} \frac{\rho_0'}{\rho_0} - k_y^2}.
\eq
In order to apply WKB without sacrifice to quantitative accuracy, we had to be warranted in assuming that once we recast our equation in order to obtain wave solutions, we could treat $C(r)$ as constant over length scales on which the wave packet varied. That is in addition to the assumed $k_r^{-1} \ll r$ and $k_r^{-1} \ll L$, where $L$ was the length scale of magnetic field and pressure gradients, which is implicit in the reduced wedge wave equation.

We can see that the first term under the square root, however, that the wave packet will vary on the scale of the density gradient. That is, we will always have $k_r \sim \rho_0' / \rho_0$. On the one hand, this unfortunately entails that errors in the WKB analysis will be of order unity, but it also immediately yields the scale on which variations in the wave packet occur! The WKB analysis will only allow for a qualitative description of the time evolution of the instability in the linear regime, but the spatial grid size for a full nonlinear analysis is now determined. Once the location of the most unstable mode in the geotail is determined, the initial instability will vary on length scales approximately equal to the width of density gradient length scales in the region.

\section{Conclusions}

In summary, the shear flow-interchange wave equation best accommodates the circumstances under which an instability in the magnetotail initiates transmission of $\vec{E} \times \vec{B}$ sheared zonal flows along magnetic flux tubes towards the E-layer of the Earth's ionosphere. WKB analysis yields a qualitative description of the most unstable propagating wave packet. A propagating wave packet undergoes growth as it travels, and is analyzed to the fullest extent that the linear analysis will allow, yielding results which can be supplemented by background parameter models to yield growth rates and dispersion properties for comparison with the auroral bead patterns mapped back along magnetic field lines to the magnetotail. For a quantitatively accurate picture, however, the full nonlinear analysis needs to be done and the reduced wedge wave equation needs solved numerically. The applicability conditions yield the spatial scale for variations in the instability for the nonlinear analysis.

The wedge model has several oversimplifications worth mentioning. One discussed above is the absence of a conductance (field-aligned current) on the angular boundaries of the wedge. Gravity and a background centrifugal acceleration were taken to be negligible. Frozen-in flux (via flux tube densities and obedience to Ohm's law) and adiabaticity were also both assumed. The model is also symmetric about the equatorial plane. As a fluid model, it neglects kinetic effects, so that relevant physical features with spatial extent smaller than the ion Larmor radius are not taken into account. There may be some limiting procedures and relations between the instabilities arising from the MHD model in this paper and the results discussed in \citeA{KAL18}, which claims that auroral beads are likely the signature of kinetic shear Alfv{\'e}n waves driven unstable in the magnetotail prior to substorm onset.

The wedge model and resulting shear flow-interchange instability analysis can be used to validate background parameter models and potentially allow for real-time prediction of substorm onset. Alternatively, obtainment of data yielding the background parameters as a function of radius in the region can be used to determine where precisely within the near-Earth magnetotail plasma sheet the instability is most likely to occur, and how an initial instability of local density gradient spatial scale size will propagate and disperse. Further work will involve modeling the full nonlinear equations using TAE integrated with background parameters from models such as the Rice Convection Model \cite{SAZ81, HAR81, WOLF83, TOFF03, WOLF07, WU09} to model the time-evolution of pressure, density, magnetic field strength, and fluid velocity in the magnetotail during magnetic substorms. The fully nonlinear wedge model will then be integrated with the WINDMI model \cite{HOR98-1, HOR98-2, HOR02, HOR05, SPE06} to attempt the real-time prediction of substorm onset and evolution.

\acknowledgments
We would like to thank Dr. Boris Breizman for the indispensable insight he provided in discussions pertaining to this research.

\bibliography{BibTex}

\appendix

\section{Voronkov Treatment of Momentum Equation}

Here I will briefly discuss the way in which the derivation of \citeA{VOR97} results from inconsistencies due to a misleading grouping of terms in Voronkov's treatment of the momentum equation.

Equilibrium force balance can be written in general as:
\bq
	- \frac{\p}{\p r} \left( P_0 + \frac{B_0^2}{2\mu_0} \right) + \frac{B_0^2}{\mu_0 r} = \frac{\rho_0 v_\phi^2}{r}.
\eq
Voronkov recasts this in the following way:
\bq
	\rho_0 g = \frac{\p}{\p r} \left( P_0 + \frac{B_0^2}{2\mu_0} \right),
\eq
, where:
\bq
	\rho_0 g := \frac{B_0^2}{\mu_0 r} - \frac{\rho_0 v_\phi^2}{r}.
\eq
Notice that $g$ contains both a force term and an acceleration term.

Subsequently, the momentum equation gets perturbed with a $\rho g$ term acting as a source, rather than $\vec{B} \cdot \nabla \vec{B}/ \mu_0$. This is not a gravitational term, but a term with the $g$ defined implicitly as above. There is no obvious reason that $g$ in this form should be treated either as a background constant (it has radial dependence) or unperturbed. As far as the perturbation goes, the magnetic field term gets perturbed everywhere except within the $\rho g$ term. Also, equilibrium quantities are said to be in agreement with ours, contradicting the previous assumption of an azimuthal component of the background velocity. It is unclear how to fully characterize the discrepancy, but it is clear that the perturbation of the momentum equation is performed incorrectly.

Recast in the above notation, the \citeA{VOR97} result (obtained in a methodologically similar way corresponding (roughly) to limits taken in Section 5, though with the additional assumption that density gradient scales are large) utilized by \citeA{KAL15} is:
\bq
	\delta v_r'' + \left( \frac{v_0''}{k_y \widetilde{\omega}} - \frac{g}{\widetilde{\omega}^2} \frac{\rho_0'}{\rho_0} - \frac{1}{\widetilde{\omega}^2} \frac{g^2}{c_f^2} - 1 \right) k_y^2 \delta v_r = 0.
\eq
As written by \citeA{VOR97}, this has the form:
\bq
	\delta v_r'' + \left( \frac{v_0''}{k_y \widetilde{\omega}} + \frac{W}{\widetilde{\omega}^2} - 1 \right) k_y^2 \delta v_r = 0,
\eq
where:
\bq
	W := - \frac{g \rho_0'}{\rho_0} - \frac{g^2}{c_f^2},
\eq
with effective acceleration defined above. The term W obtained by \citeA{VOR97} was taken to be an analog of the buoyancy frequency, $\omega_b^2$. Near the inner edge of the plasma sheet, at approximately 9-12 $\, R_E$, the analysis of \citeA{KAL15} indicates that these terms are destabilizing, whereas the above analysis reveals these terms to be stabilizing. This is due to differences between ballooning and interchange, where the former are often treated as localized and the latter is globally distributed along the magnetic field lines. The interchange instability results, rather than ballooning, since the unstable mode does not perturb the background equilibrium magnetic field.

Interchange instability is just a special case of the ballooning instability in which the unstable mode does not perturb the magnetic field lines. It is possible (though not necessary) that relaxing constraints on the wedge model will provide a real source of ballooning modes which would replace the interchange modes. But in order to obtain actual ballooning modes, the magnetic field lines must be free to move, which they are unable to do under the \citeA{VOR97} analysis. This should have been a phenomenological hint to the incorrectness of the result. The main point to stress is that it is the equation which is incorrect, but that a more correct treatment may indeed replace the shear flow-interchange instability with a shear flow-ballooning instability of some sort.

\section{Interchange Instability Analysis}

It should be mentioned that the limits from Section 5, taken in the appropriate order, agree with those in \citeA{WOLF18}, barring what appear to be minor typographical errors (as seen by a unit check) on his part. The buoyancy frequency, which was thoroughly discussed in \citeA{WOLF18}, is given by the next-to-last term in our equation. Let us perform this check. Dropping the shear velocity terms in (\ref{SFI}), we obtain:
\bq
	\delta v_r'' + \frac{\rho_0'}{\rho_0} \delta v_r' + \left( \frac{1}{\omega^2} \frac{2}{\Gamma r} \frac{c_A^2 c_s^2}{c_f^2} \frac{K_0'}{K_0} - 1 \right) k_y^2 \delta v_r = 0
\eq
Thus the first term in parentheses yields immediately the buoyancy frequency for waves in a wedge:
\bq
	\omega_b^2(r) = \frac{2}{\Gamma r} \frac{c_A^2 c_s^2}{c_f^2} \frac{K_0'}{K_0},
\eq
where $K$ is the entropy, defined:
\bq
	K(r) := P V^\Gamma = \frac{P}{\rho^\Gamma}
\eq
It is demonstrated in \citeA{WOLF18} that this is just the oscillation frequency of thin magnetic filaments in the wedge. The speed $c_A c_s/c_f$ is just that of the slow mode buoyancy waves which result from pure interchange oscillations. Notice also that the frequencies are no longer Doppler-shifted, as there is no fluid velocity to supply the shift.

The interchange dispersion relation from \citeA{WOLF18}, absent shear flow and density gradients is given by:
\bq
	\omega^2 - \frac{k_y^2}{k_y^2 + k_r^2} \frac{2}{\Gamma r} \frac{c_A^2 c_s^2}{c_f^2} \frac{K_0'}{K_0} = 0.
\eq
By examination, we can see that there is no oscillatory growth. There is either propagation in the absence of growth or growth in the absence of propagation, and the condition for growth is just that:
\bq
	\frac{K_0'}{K_0} < 0.
\eq
This situation is analyzed in much detail in \citeA{WOLF18}.

\end{document}